\documentclass[%
 preprint,prl,
showpacs,
 amsmath,amssymb,
 aps
]{revtex4-1}

\usepackage{amsmath}
\usepackage{amssymb}
\usepackage{graphicx}
\usepackage{dcolumn}
\usepackage{bm}
\usepackage{hyperref}
\usepackage{verbatim}
\usepackage{color}

\newcommand{\vrva}{\mathcal{\nu}}

\renewcommand{\vec}{\textbf}	

\newcommand{\CH}[1]{\textcolor{black}{{#1}}}

\begin{document}

\title{Defect-Mediated Phase Transitions in Active Soft Matter}

\author{Christoph A. Weber}
\author{Christopher Bock}
\author{Erwin Frey}
\affiliation{%
Arnold Sommerfeld Center for Theoretical Physics and Center for NanoScience, \\
Department of Physics, Ludwig-Maximilians-Universit\"at M\"unchen, Theresienstra\ss e 37, D-80333 Munich, Germany
}

\begin{abstract}
How do topological defects affect the degree of order in active matter? 
To answer this question we investigate an agent-based model of self-propelled particles, which accounts for polar alignment and short-ranged repulsive interactions.  For strong alignment forces we find collectively moving polycrystalline states with  fluctuating networks of grain boundaries. In the regime where repulsive forces dominate, the fluctuations generated by the active system give rise to quasi-long-range transitional order, but---unlike thermal system---without creating topological defects. 
\end{abstract}



\maketitle

For a system in thermodynamic equilibrium, phases with a broken continuous symmetry in two spatial dimensions are prohibited by general theorems~\cite{Mermin_Wagner:66, Hohenberg:67}. Yet, for two-dimensional solids, XY magnets, and superfluids there is a clear qualitative difference between a low-temperature phase exhibiting quasi-long-range order and a high-temperature phase where correlation functions decay exponentially~\cite{Nelson:Book}. 
\CH{Since for crystalline solids the low- and high-temperature phases are separated by two broken symmetries, namely translational and orientational symmetry, melting can proceed by two steps~\cite{Kosterlitz_Thouless:73, Young:79, Halperin_Nelson:78, Nelson_Halperin:79}: The unbinding of dislocation pairs drives a continuous phase transition from a crystalline phase with quasi-long-range translational order into a hexatic phase~\cite{Halperin_Nelson:78, Nelson_Halperin:79} with remaining quasi-long-range bond orientational order. This is followed by another continuous transition into a disordered liquid phase mediated by the proliferation of isolated disclinations.}

These statements may no longer remain valid for systems driven out of thermodynamic equilibrium. 
Indeed, for active systems where individual particles are self-propelled, an antagonism between dissipative processes favoring ``ferromagnetic'' alignment of the particles' velocities and noise can trigger a phase transition from an isotropic to a long-range ordered polar state, where particles move collectively. This was first demonstrated by Vicsek et al.~\cite{Vicsek} who employed a two-dimensional agent-based model where particle alignment is implemented as an update rule: Each particle aligns parallel to the average of all particles' orientations within some defined finite neighborhood. Interestingly, computer simulations of the Viscek model show that the transition is discontinuous, and the polar state exhibits propagating wave-like excitations~\cite{Chate:2004,Chate:2008}. The basic mechanism for the phase transition is believed to constitute a low-density phenomenon that is amenable to a kinetic description~\cite{Bertin_short:2006,Bertin_long:2009,Weber_NJP_2013, Thueroff_Weber_Frey_2013,Hanke_Weber_Frey_2013}. It assumes that the formation of order is driven by a gradual reduction in the spread of particle orientations by means of weakly aligning binary collisions. 
Experimental investigations supporting this picture are motility assays where cytoskeletal filaments are propelled \CH{by} a lawn of molecular motors~\cite{Butt,Schaller,Schaller2,Yutaka}, and vibrated granular systems~\cite{Dauchot_Chate_2010,Weber:2013}.

In contrast, much less is known about ordered states of active matter at \emph{high densities}, where in addition to \emph{polar order} the active system may also exhibit different degrees of \emph{liquid crystalline}~\cite{Sanchez:2012} or even \emph{crystalline order}. Numerical studies of models for (self-)propelled particles discovered jammed~\cite{Henkes:2011}, and also crystalline-like states at large packing fractions~\cite{Chate:2004, Bialke:12}. Recently, a mean-field theory combining elements from phase-field models of crystals~\cite{Elder:2002} and hydrodynamic theories of active systems~\cite{Toner_Tu_1995,Toner_Tu_1998, Toner:2012} was proposed and shown to exhibit a wealth of crystalline states of different symmetry and degrees of polar order~\cite{Menzel:2013}. Although all these theoretical studies suggest the interesting possibility of the emergence of translational and orientational order in active particle systems, a characterization of the nature of these ordered states and the transition between them remains elusive. In this context, one might suppose that topological defects will play an important role. Indeed, recent experimental and theoretical studies of active liquid crystals~\cite{Sanchez:2012, Giomi:2013,Pismen_2013} show that activity leads to generation and swarming of topological defects. Active dislocations have also been shown to drive growth of bacterial cell walls through dislocation climb~\cite{Amir:2012}.

Here we investigate the role of topological defects for the nature of ordered states in active matter at high densities. To this end, we build on a generalized Vicsek model introduced by Gr\'egoire and Chat\'e~\cite{Gregoire:2003, Chate:2004}, which accounts for alignment as well as short-ranged repulsive interactions. Depending on their relative strength we find different degrees of crystalline and polar order. In the repusion-dominated regime we find no polar order, \emph{i.e.}\ no collective motion of the particles. Interestingly, however, fluctuations generated by the active system lead to an intriguing state of matter exhibiting quasi-long-range translational order but---unlike systems in thermal equilibrium---devoid of any topological defects. In contrast, in the regime where dissipative alignment dominates we find collectively moving polycrystalline states with hexagonally ordered crystalline domains of characteristic size. These states exhibit pronounced defect fluctuations and sound-wave-like excitations. 

To study active soft matter at high densities, we consider an off-lattice system of $N$ particles which have a tendency to align their velocity with neighboring particles and repel each other if they come too close~\cite{Gregoire:2003,Chate:2004}. These interactions are implemented by the following parallel update rules for the velocity $\vec{v}_i(t)$ and position $\vec{x}_i(t)$ of each particle $i$ with some discrete time interval $\Delta t$:
\begin{eqnarray}
\label{eq:velupdate}
\vec{v}_i(t+\Delta t) &=& v_a\frac{\sum_{j \in \mathcal{A}_i}^{}  \vec{n}_j(t) }{|\sum_{j \in \mathcal{A}_i}^{}  \vec{n}_j(t)|} + v_r {\sum_{j \in \mathcal{A}_i}}^\prime \frac{\vec{x}_{ij}(t)}{|\vec x_{ij}(t)|} \, , \\
\vec{x}_i(t+\Delta t)&=&\vec{x}_i\left(t\right) + \vec{v}_i\left(t+\Delta{}t\right)\Delta{}t \, .
\label{eq:posupdate}
\end{eqnarray}
Here $\vec{n}_i := {\vec{v}_i}/{\left|\vec{v}_i\right|}$ denotes the particle director, and $\vec{x}_{ij} := {\vec{x}_i  - \vec{x}_j}$ signifies the relative position vector between particles $i$ and $j$. The first term in Eq.~\eqref{eq:velupdate} is an alignment interaction as introduced by Vicsek et al.~\cite{Vicsek} where the updated velocity of particle $i$ is given by the average velocity of all particles within a circular area $\mathcal{A}_i$ of radius $2R$ centered on particle $i$. The parameter $v_a$ characterizes the strength of alignment as well as the particles' propulsion speed. The second term in Eq.~\eqref{eq:velupdate} describes a soft,  pairwise additive repulsive interaction between a given particle $i$ and all its neighbors within the same area  $\mathcal{A}_i$. It displaces a particle pair, whose separation $|\vec x_{ij}(t)| \le 2R$ [$i\not=j$, indicated by the primed sum], radially outward by a constant amount $v_r \, \Delta t$. In the following, we will refer to $R$ as the particle radius. Length and time are measured in units of the particle diameter, $2R$, and the corresponding time to traverse this distance, $\tau = {2R}/{v_a}$, respectively. The model can easily be generalized to account for different radii for alignment and repulsion, $R_a$ and $R_r$, respectively. Here, we focus on the competition between alignment and repulsion, and therefore have chosen the two radii as $2R_r=R_a$; previous studies of the Vicsek model with repulsion were restricted to the limit $R_r \ll R_a$~\cite{Chate_Variations,peng_vicsek_rep_adaptive_speed}. We are mainly interested in the collective dynamics as a function of the packing fraction $\rho=N \, \pi R^2/L^2$, and the relative strength of the repulsive and alignment interaction $\nu:=v_r/v_a$.

First, we analyze the degree of polar and bond-orientational order. The global polarization is defined as a system average $\langle...\rangle_i$ over all particle orientations $\vec{n}_i(t)$: $\mathcal{P}(t)=|\langle\vec{n}_i(t)\rangle_i|$. Local bond-orientational order is characterized by the hexatic order parameter $\Psi_{6,i}=|\mathcal{N}_i|^{-1} \sum_{j\in{}\mathcal{N}_i}e^{ \imath 6\theta_{ij}}$, where summation extends over all $\mathcal{N}_i$ topological (Voronoi) nearest neighbors of particle $i$, and $\theta_{ij}$ is the `bond'-angle between particles $i$ and $j$ relative to an arbitrarily chosen reference axis.

\begin{figure}[tb]
\includegraphics[width=\linewidth]{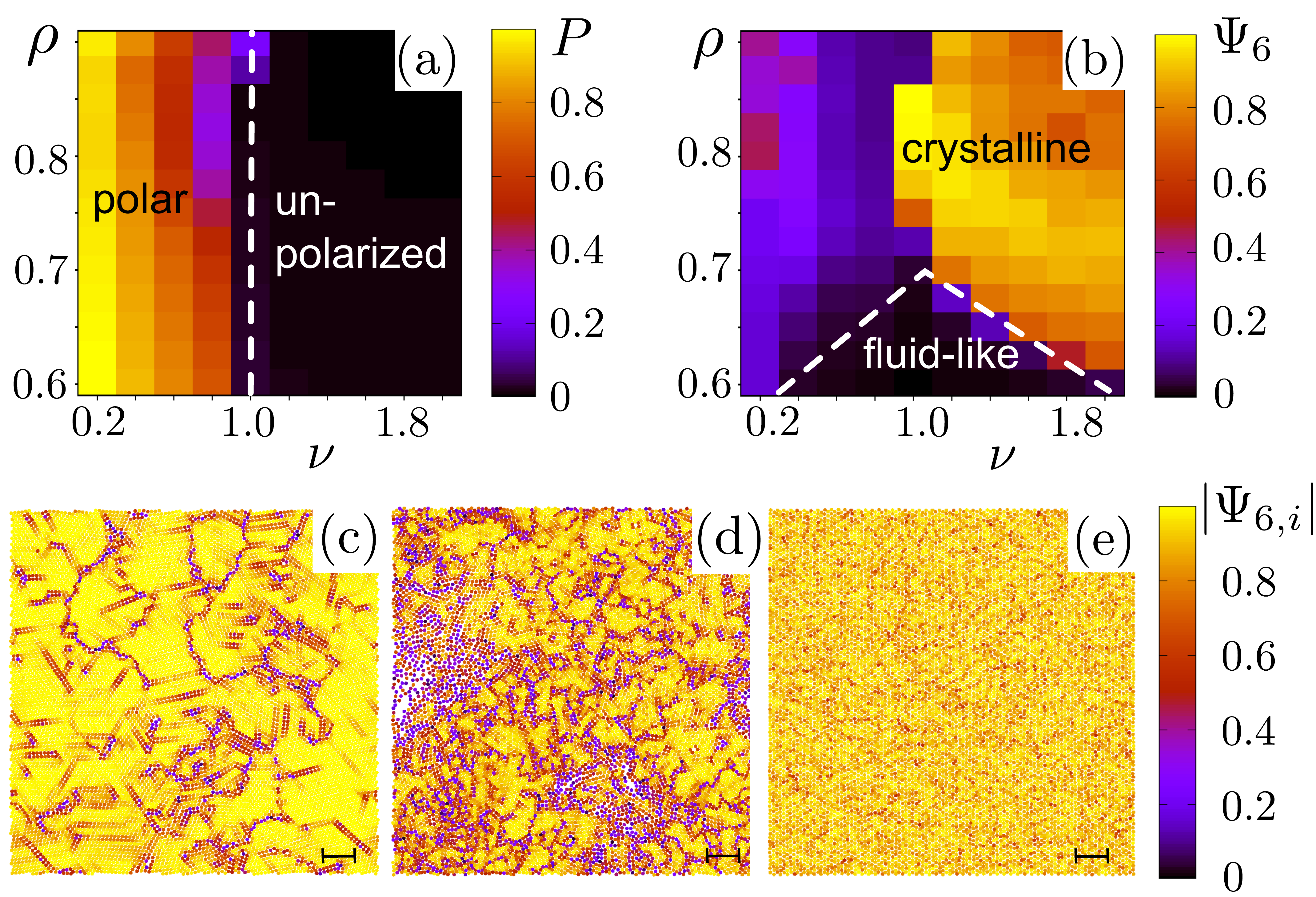}
\caption{\label{fig:phasediagram}
Global polar order parameter $P$ \textbf{(a)} and hexatic order parameter $\Psi_6$ \textbf{(b)}
as a function of the control parameters $\rho$ and $\nu$.  The dashed white lines indicate tentative boundaries between \emph{polar} and \emph{unpolarized} states, and \emph{crystalline} states exhibiting crystalline order from \emph{fluid-like} states. 
Snapshots of local hexatic order $|\Psi_{6,i}|$ for a relative interaction value $\nu=0.25$  \textbf{(c)},  $\nu=0.75$  \textbf{(d)},   and  $\nu=1.5$ \textbf{(e)}; all three snapshots correspond to a high packing fraction of $\rho=0.85$. See also videos in the Supplemental Material~\cite{SM}. 
Scale bars indicate a distance of $20R$.
}
\end{figure}

Figures~\ref{fig:phasediagram}(a,b) illustrate the degree of polar and hexatic order as a function of the packing fraction $\rho$, and the relative strength of repulsive and alignment interactions $\nu$. We observe that global polar order, characterized by the time-averaged polarization $P=\langle \mathcal{P}(t) \rangle_t$ [$\langle...\rangle_t$: time-average], is well-established for weak repulsion $\nu \lesssim 1$, but at $\nu \approx 1$  sharply drops to very small values [Fig.~\ref{fig:phasediagram}(a)]. The respective phase boundary between \emph{polar} and \emph{unpolarized} states is tentatively defined by $P=0.2$. Note that it is nearly independent of the packing fraction $\rho$, indicating that the transition from a polar collectively moving state to an unpolarized state is mainly driven by an antagonism between repulsive and alignment forces but not the particle density. To discern the different degrees of bond-orientational or translational order is more difficult. As can be inferred from Fig.~\ref{fig:phasediagram}(b), there are different degrees of global hexatic order, $\Psi_6=\langle \left|\langle \Psi_{6,i} \rangle_i \right|\rangle_t$, with a maximum for large packing fraction $\rho$ and strong repulsive interaction (large $\nu$); the dashed white line in Fig.~\ref{fig:phasediagram}(b) correspond to a value of  $\Psi_6= 0.2$. Strikingly, the loss of polar order is concomitant with the emergence of a high degree of crystalline order, and vice versa.

\begin{figure}[t]
\includegraphics[width=0.8\linewidth]{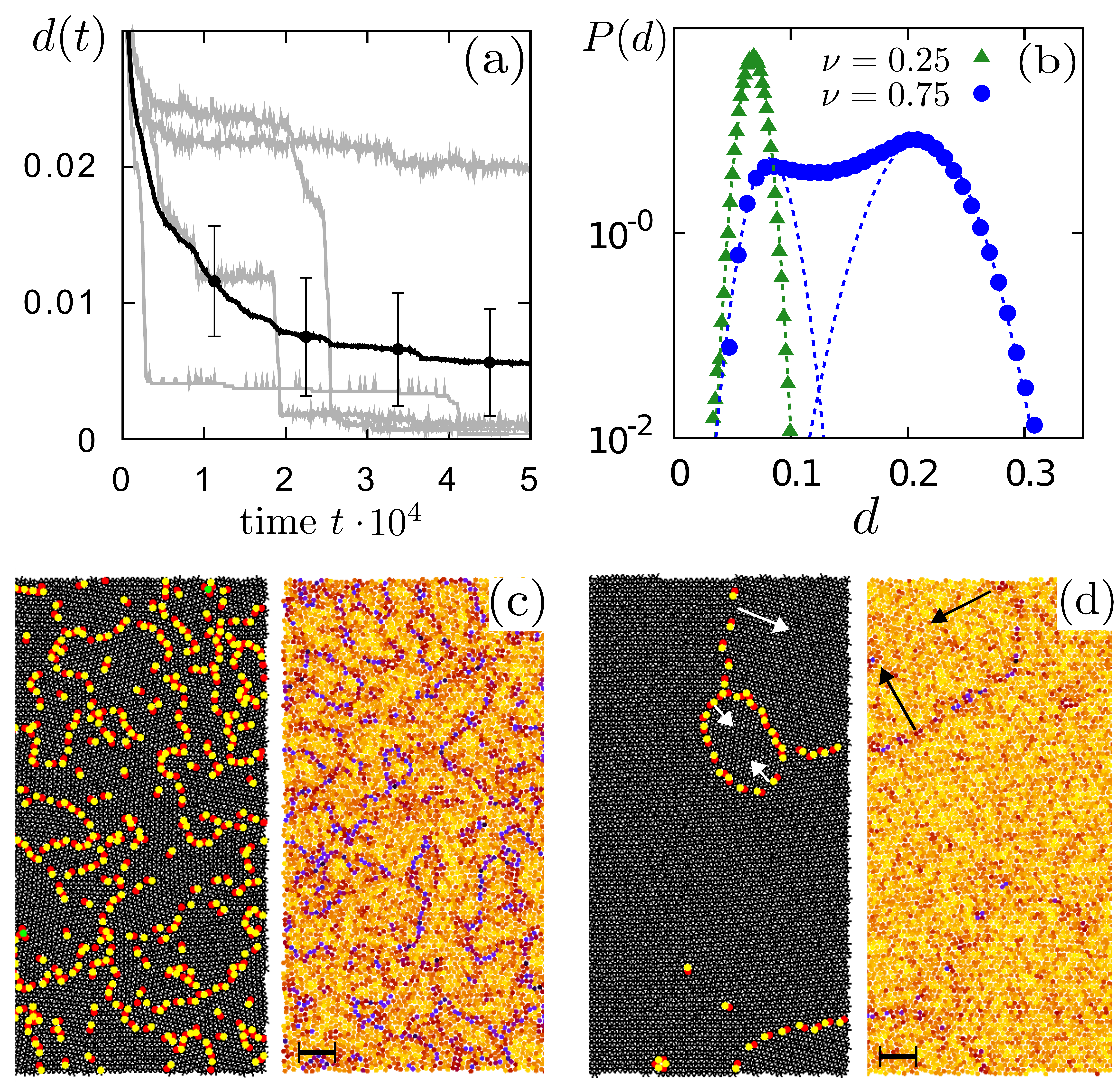}
\caption{\label{fig:defectCount}
\textbf{(a)} Defect ratio $d(t)$ as a function of time $t$ for  $(\rho, \nu) = (0.85, 1.5)$. The black curve is an average over $50$ realizations (grey lines) with error bars indicating the standard deviation. 
The steep decline in the time traces  corresponds to the fast annihilation processes of grain boundaries [see \textbf{(d)}, white errors].
\textbf{(b)}  Probability distribution $P(d)$ [lin(d)-log] of the defect ratio for two stationary (flowing) polycrystalline states, i.e.\ $(\rho,\nu)=(0.85, 0.25)$ (green, triangles) and $(\rho,\nu)=(0.85, 0.75)$ (blue, circles). 
\textbf{(c,d)} Snapshots illustrating the defect dynamics for an active crystal [$(\rho,\nu)=(0.85,1.5)$], where the system shows a high degree of crystalline order: \textbf{(c)} Early phase with roughly homogeneous distributed defects and small hexagonal patches; \textbf{(d)} Formation of ring-like grain boundaries at larger times that contract (indicated by black arrows), leading to a sudden decrease of $d(t)$ [see \textbf{(a)}].  The left and right half of each figure depict the \emph{Voronoi} triangulation and the local hexatic order parameter $|\Psi_{6,i}|$, respectively. Disclinations are indicated by red/yellow dots, and green dots corresponds to particles with more than $7$-fold or less than $5$-fold coordination. Scale bars: $20R$.
}
\end{figure}

While the global polar and hexatic order parameters provide a first rough estimate of the degree and nature of the ordered states, a full characterization thereof requires an in-depth analysis of the spatio-temporal dynamics. In particular, as for thermodynamic equilibrium systems, the dynamics and the spatial organization of topological defects are especially important indicators of crystalline order. Figs.~\ref{fig:phasediagram}(c-e) depict snapshots of the local hexatic order $\left|\Psi_{6,i}\right|$ at a large packing fraction of $\rho=0.85$ for a set of values for $\nu$. Depending on the relative strength of repulsive and alignment interaction marked differences in the spatial organization of defects are clearly visible. While for $\nu \lesssim 0.25$ dislocations
align to form a network of rather well-defined grain boundaries, they tend to cluster in the intermediate regime $0.375 \lesssim \nu \lesssim 1.0$ [Figs.~\ref{fig:phasediagram}(c,d)]. For $\nu \gtrsim 1.0$, concomitant with the loss of polar order, the defects become more evenly spread and slowly disappear from the system; see Fig.~\ref{fig:phasediagram}(e) for a snapshot, and Fig.~\ref{fig:defectCount}(a) for the dynamics of the defect density. This reassures the observation made on the basis of the order parameters, namely that polar and crystalline order are mutually exclusive.

The intricate interplay between crystalline and polar order is elucidated by the spatio-temporal dynamics of the defects; see the videos in the Supplemental Material~\cite{SM}. For $\nu \lesssim 0.375$, we observe a \emph{flowing polycrystalline state} where changes in the flow direction strongly affect the network of grain boundaries. In the stationary regime, the defect fraction $d = D/N$ is Gaussian-distributed around a mean of about $7\%$ [Fig.~\ref{fig:defectCount}(b)];  here $D$ is the number of all particles with a coordination different from $6$-fold. In the intermediate regime, we find \emph{intermittent dynamics} where episodes of polycrystalline and polar order alternate with episodes of disorder which are accompanied by sound waves (see Supplemental Material~\cite{SM}, S1, S4). These compression waves, caused by collisions of polar crystalline domains, impede their growth into larger domains and thereby rule out the coexistence of polar and crystalline order. Moreover, the intermittent dynamics is also reflected in a bimodal shape of the defect probability density $P(d)$ [Fig.~\ref{fig:defectCount}(b)]: While the peak at low values of $d$ corresponds to particle configurations with a high degree of polar order, the peak at higher values originates from time intervals where collective motion breaks down and strong density inhomogeneities arise.

Dynamics and spatial organization of topological defects change qualitatively for strong repulsive interaction, $\nu \gtrsim 1$, where  polar order is also absent. Starting from an initial disordered state [Fig.~\ref{fig:defectCount}(c)], we observe that first the spatial distribution of defects  coarsens quickly and then organizes into grain boundaries [Fig.~\ref{fig:defectCount}(d)]. Subsequently these grain boundaries contract and self-annihilate, leaving the system in a state with evenly spread disclinations and dislocations; see also videos in the Supplemental Material~\cite{SM}. The ring-like annihilation processes of these grain boundaries 
 are seen as periods of steep decline in the time traces for the defect fraction [Fig.~\ref{fig:defectCount}(a), grey curves]. After each steep decline, the decrease in defect number slows down significantly due to an enlarged inter-defect distance. We observe that the number of isolated defects decreases extremely slowly; see the asymptotic decline in the average defect fraction in Fig.~\ref{fig:defectCount}(a). Moreover, we find evidence that the topological defects even move sub-diffusively (see Supplemental Material~\cite{SM}, S3). Taken together, it is numerically not feasible to study the asymptotic dynamics significantly beyond what is shown in Fig.~\ref{fig:defectCount}(a). To check whether a defect-free crystal is stable we initialized the system in an unpolarized and perfectly hexagonal ordered state, and waited until the global hexatic order parameter $\Psi_6(t)$ converged to a stationary value. Even though the active dynamics leads to a reduction of the hexatic order parameter to a stationary value of $\Psi_6 \approx 0.9$, it is not strong enough to create any defects for densities larger than $\rho\approx0.8$. Hence, we conclude that the stationary states for $\nu \gtrsim 1$ and large density ($\rho\gtrsim0.8$) are indeed \emph{free of topological defects}. When decreasing the packing fraction below $0.8$, there is a small range of packing fractions where fluctuations trigger the creation of defects (see Supplemental Material~\cite{SM}, S2). However, in the ensuing non-equilibrium steady states corresponding to this smallish transitional region, dislocations 
 are always found in pairs. As in case of thermal systems this would indicate that the corresponding states also exhibit quasi-long-range order.
  Exploring parameter space we could not identify hexatic phases with isolated dislocations.
  Further decreasing the packing fraction, the defect ratio $d(\rho)$ increases to a rather high value $d \sim 0.4$, signaling a transition to a fluid-like phase. 

In order to further scrutinize the nature of order within the crystalline regime we computed the pair correlation function $g(\vec r)$, the corresponding static structure factor  $S(\vec q)$, and the correlation functions~\cite{Nelson:Book, SM} 
\begin{equation}
C_\alpha (r)= \frac{1}{\sum_{i}  |\Psi_{\alpha, i}|^2} \sum_{|\vec x_i-\vec x_j|=r}  \Psi_{\alpha, i} \Psi^*_{\alpha, j} \, 
\end{equation}
for the hexatic $\Psi_{6,i}$, and the translational $\Psi_{\vec{G},i}=e^{- \imath \, \vec{G} \, \vec{r}_i}$ order parameter with $\vec{G}$ denoting a reciprocal lattice vector. As discussed above, for $\nu>1$, dislocations vanish extremely slowly and, as a consequence, the asymptotic non-equilibrium steady state cannot be reached within a computationally accessible time. Therefore, to obtain steady state results for the correlation functions, we initialized  the system in a hexagonal and isotropic configuration ($\Psi_6=1$, $P\approx0$); the corresponding results for simulations starting from an disordered initial state are discussed in the Supplemental Material~\cite{SM}. We find that both $S(\vec q)$ [Figs.~\ref{fig:ffts2d}(a) and \ref{fig:ffts2d}(b)] and $g(\vec r)$ [see Supplemental Material S5] exhibit a sharp and discrete pattern of hexagonal symmetry, clearly indicating a high degree of translational order. This is confirmed by $C_6 (r)$ being constant over the whole system size [Fig.~\ref{fig:ffts2d}(c)], and the slow decay of the translational correlation function $C_{\vec{G}}(r)$. The decay follows a power-law with a very small exponent of about $0.04$  [see Fig.~\ref{fig:ffts2d}(d), dashed grey line], which is difficult to discern from a logarithmic decay. Taken together, these results lead us to conclude that this state of active matter is an \emph{active crystal}, free of topological defects with long-range bond-orientational order and quasi-long-range translational order. In contrast, in the parameter regime of \emph{polycrystalline} order,  the static structure factor shows the ring-like features of a liquid [Figs.~\ref{fig:ffts2d}(b)]; see also Supplemental Material~\cite{SM} S5. These features are due to the different orientations of the hexagonally ordered patches, as also evident from the exponential decay in $C_6$ [Fig.~\ref{fig:ffts2d}(c)] and the fast decay of $C_{\vec{G}}(r)$ [Figs.~\ref{fig:ffts2d}(e)].

\begin{figure}[t]
\includegraphics[width=\linewidth]{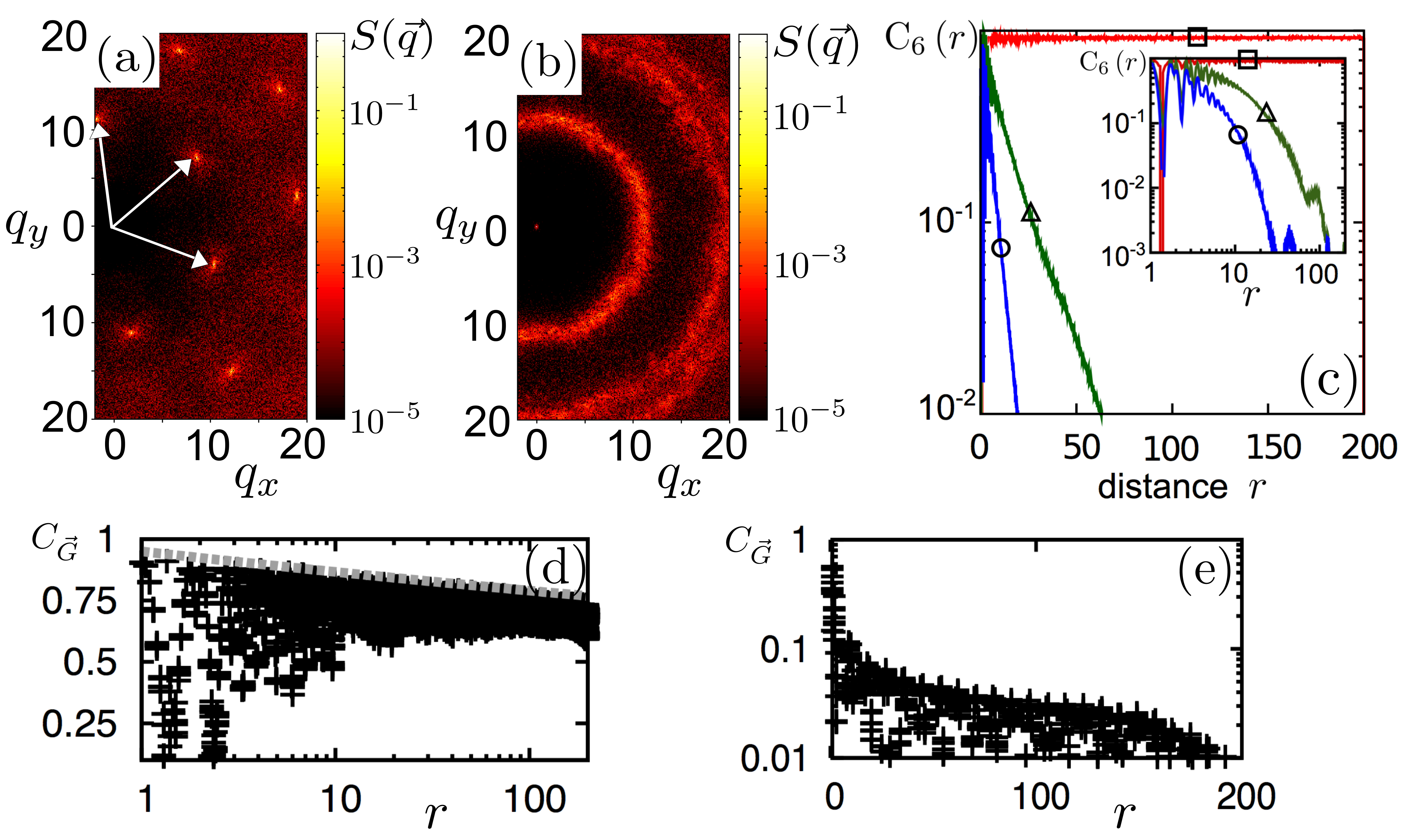}
\caption{\label{fig:ffts2d} 
\emph{Static structure factor $S(\vec q)$} for \textbf{(a)}~the active crystal [$(\rho,\nu)=(0.85,1.5)$]  and \textbf{(b)}~a polycrystal [$(\rho,\nu)=(0.85, 0.25)$], respectively, both in the stationary regime. Reciprocal lattice vectors $\vec G$ are indicated by white arrows. 
\textbf{(c)} \emph{Correlation function $\text{C}_6(r)$}
 in lin(r)-log (\emph{inset:} log-log)  for the same parameters as in Figs.~\protect{\ref{fig:phasediagram}(c-d)}: $\rho=0.85$, and $\nu=1.5$ (red, square),  $\nu=0.75$ (blue, circle), and $\nu=0.25$ (green, triangle). 
\emph{Correlation function $C_{\vec{G}}(r)$} for $(\rho,\nu)=(0.85,1.5)$~\textbf{(d)} [log(r)-lin, dashed line is a power law with exponent $0.04$] and $(\rho,\nu)=(0.85, 0.25)$~\textbf{(e)}. All results correspond to a simulation box of size $L=400$ containing $N=172156$ particles; see Supplemental Material~\cite{SM} for on the data evaluation. 
}
\end{figure}


Topological defects are the hallmark of phase transitions in two-dimensional crystalline systems. For systems in thermodynamic equilibrium they drive the successive breaking of translational and bond-orientational order. Our investigations of active crystalline matter at high density have revealed: While defects still play a decisive role, the emerging defect dynamics and phase behavior differ qualitatively from their equilibrium analogues. In active systems, the non-equilibrium steady states include different types of polycrystalline phases, and a crystalline phase with quasi-long-range translational order but completely devoid of any topological defects.  
How the genuine differences of 
the fluctuations generated by the active particle motion compared to thermal fluctuations permit a defect-free state remains presently unclear and constitutes an interesting future challenge for a continuous theory describing the defect dynamics in polar active matter. One possibility is to develop a Langevin description for the defect dynamics including interaction potentials that can be directly measured in the simulations. In regimes where defects are sparse, an alternative promising route might be to employ kinetic approaches~\cite{Bertin_short:2006,Bertin_long:2009, Thueroff_Weber_Frey_2013,Hanke_Weber_Frey_2013}. 

Our predictions can readily be tested by experimental model systems which may for instance be realized using emulsion droplets containing extensile microtubule bundles~\cite{Sanchez:2012}. They exhibit spontaneous motility when in frictional contact with a hard surface; depending on the availability of ATP their motion can be tuned from passive Brownian motion to active persistent random walks. We envisage that large assemblies of such active soft droplets are ideal model systems to test our theoretical predictions. Though the detailed mechanisms of the interaction between the droplets is different from the interaction rules of the agent-based model, we expect that the main features of the dynamics and phase behavior to be generic for active matter at high densities. Another promising experimental system are active colloidal particles. Recent studies of photo-activated colloidal particles~\cite{Palacci_Chaikin_2012} and carbon-coated Janus particles~\cite{Buttinoni_Speck_2013} show various types of pattern and cluster formation. The versatility of colloidal systems should also allow the design of experiments to explore the dynamics of active matter at high density.

\begin{acknowledgments}
We would like to thank David Nelson for fruitful and stimulating discussions. This project was supported by the Deutsche Forschungsgemeinschaft in the framework of the SFB 863, and the German Excellence Initiative via the program ``NanoSystems Initiative Munich'' (NIM).  
\end{acknowledgments}

\newpage
\cleardoublepage


\begin{thebibliography}{10}

\bibitem{Mermin_Wagner:66}
N.~D. Mermin and H. Wagner, Phys. Rev. Lett. {\bf 17},  1133  (1966).

\bibitem{Hohenberg:67}
P. Hohenberg, Physical Review {\bf 158},  383  (1967).

\bibitem{Nelson:Book}
D.~R. Nelson, {\em {Defects and Geometry in Condensed Matter Physics}}
  (Cambridge University Press, Cambridge, 2002), p.\ 392.

\bibitem{Kosterlitz_Thouless:73}
J.~M. Kosterlitz and D.~J. Thouless, Journal of Physics C: Solid State Physics
  {\bf 6},  1181  (1973).

\bibitem{Young:79}
A.~P. Young, Phys. Rev. B {\bf 19},  1855  (1979).

\bibitem{Halperin_Nelson:78}
B.~I. Halperin and D.~R. Nelson, Phys. Rev. Lett. {\bf 41},  121  (1978).

\bibitem{Nelson_Halperin:79}
D.~R. Nelson and B.~I. Halperin, Phys. Rev. B {\bf 19},  2457  (1979).

\bibitem{Vicsek}
T. Vicsek {\it et~al.}, Phys. Rev. Lett. {\bf 75},  1226  (1995).

\bibitem{Chate:2004}
G. Gr\'egoire and H. Chat\'e, Phys. Rev. Lett. {\bf 92},  025702  (2004).

\bibitem{Chate:2008}
H. Chat\'e, F. Ginelli, G. Gr\'egoire, and F. Raynaud, Phys. Rev. E {\bf 77},
  046113  (2008).

\bibitem{Bertin_short:2006}
E. Bertin, M. Droz, and G. Gr\'egoire, Phys. Rev. E {\bf 74},  022101  (2006).

\bibitem{Bertin_long:2009}
E. Bertin, M. Droz, and G. Gr\'egoire, Journal of Physics A: Mathematical and
  Theoretical {\bf 42},  445001  (2009).

\bibitem{Weber_NJP_2013}
C.~A. Weber, F. Th\"uroff, and E. Frey, New J. Phys. {\bf 15},  045014  (2013).

\bibitem{Thueroff_Weber_Frey_2013}
F. Th\"uroff, C.~A. Weber, and E. Frey, Phys. Rev. Lett. {\bf 111},  190601
  (2013).

\bibitem{Hanke_Weber_Frey_2013}
T. Hanke, C.~A. Weber, and E. Frey, Phys. Rev. E {\bf 88},  052309  (2013).

\bibitem{Butt}
T. Butt {\it et~al.}, Journal of Biological Chemistry {\bf 285},  4964  (2010).

\bibitem{Schaller}
V. Schaller {\it et~al.}, Nature {\bf 467},  73  (2010).

\bibitem{Schaller2}
V. Schaller, C. Weber, E. Frey, and A.~R. Bausch, Soft Matter {\bf 7},  3213
  (2011).

\bibitem{Yutaka}
Y. Sumino {\it et~al.}, Nature {\bf 483},  448  (2012).

\bibitem{Dauchot_Chate_2010}
J. Deseigne, O. Dauchot, and H. Chat\'e, Phys. Rev. Lett. {\bf 105},  098001
  (2010).

\bibitem{Weber:2013}
C.~A. Weber {\it et~al.}, Phys. Rev. Lett. {\bf 110},  208001  (2013).

\bibitem{Sanchez:2012}
T. Sanchez {\it et~al.}, Nature {\bf 491},  431  (2012).

\bibitem{Henkes:2011}
S. Henkes, Y. Fily, and M.~C. Marchetti, Phys. Rev. E {\bf 84},  040301
  (2011).

\bibitem{Bialke:12}
J. Bialk\'e, T. Speck, and H. L\"owen, Phys. Rev. Lett. {\bf 108},  168301
  (2012).

\bibitem{Elder:2002}
K.~R. Elder, M. Katakowski, M. Haataja, and M. Grant, Phys. Rev. Lett. {\bf
  88},  245701  (2002).

\bibitem{Toner_Tu_1995}
J. Toner and Y. Tu, Phys. Rev. Lett. {\bf 75},  4326  (1995).

\bibitem{Toner_Tu_1998}
J. Toner and Y. Tu, Phys. Rev. E {\bf 58},  4828  (1998).

\bibitem{Toner:2012}
J. Toner, Phys. Rev. E {\bf 86},  031918  (2012).

\bibitem{Menzel:2013}
A.~M. Menzel and H. L\"owen, Phys. Rev. Lett. {\bf 110},  055702  (2013).

\bibitem{Giomi:2013}
L. Giomi, M.~J. Bowick, X. Ma, and M.~C. Marchetti, Phys. Rev. Lett. {\bf 110},
   228101  (2013).

\bibitem{Pismen_2013}
L.~M. Pismen, Phys. Rev. E {\bf 88},  050502  (2013).

\bibitem{Amir:2012}
A. Amir and D.~R. Nelson, Proceedings of the National Academy of Sciences {\bf
  109},  9833  (2012).

\bibitem{Gregoire:2003}
G. Gr\'egoire, H. Chat\'e, and Y. Tu, Physica D: Nonlinear Phenomena {\bf 181},
   157   (2003).

\bibitem{Chate_Variations}
H. Chat\'e {\it et~al.}, The European Physical Journal B - Condensed Matter and
  Complex Systems {\bf 64},  451  (2008).

\bibitem{peng_vicsek_rep_adaptive_speed}
L. Peng {\it et~al.}, Phys. Rev. E {\bf 79},  026113  (2009).

\bibitem{SM}
See Supplemental Material for videos and more information at http://...., which includes Refs. \cite{Chate:2008,underestimation_of_trans_order,Weeks_Weitz}.

\bibitem{Palacci_Chaikin_2012}
J. Palacci {\it et~al.}, Science {\bf 339},  936  (2013).

\bibitem{Buttinoni_Speck_2013}
I. Buttinoni {\it et~al.}, Phys. Rev. Lett. {\bf 110},  238301  (2013).

\bibitem{underestimation_of_trans_order}
C.-C. Liu {\it et~al.}, Journal of Polymer Science Part B: Polymer Physics {\bf
  48},  2589  (2010).

\bibitem{Weeks_Weitz}
E.~R. Weeks and D.~A. Weitz, Phys. Rev. Lett. {\bf 89},  095704  (2002).


\end{thebibliography}

\cleardoublepage
\newpage
\begin{center}
\section{Supplemental Material:}
\end{center}
All simulations of the agent-based model were, if not stated otherwise, performed in a square box of side-length $L=100$ [unit length: particle Diameter $2R$] with periodic boundary conditions and  typically containing  $N \sim 10^4$ particles. However, the characterization of states via correlation functions (see Fig.~3, main text) is based on simulations performed in a box of side-length $L=400$ containing about $2\cdot10^5$ particles.

Noting that $v_a \, \Delta t$ sets the maximal penetration depth in  binary collisions, we choose the other parameters such that $v_a \Delta t \ll 2R$ in order to reduce the number of events where particles would pass through each other. Specifically, the updating time is fixed to $\Delta t=1$, and we take $v_a=0.05$. 

\subsection{Initialization at high densities}

In general, we initialized the numerical simulations of the agent based model in two different configurations, which we termed:
\vspace{0.1cm}\\
\emph{($\mathcal{R}$) Random.} 
Particle initial positions and orientations were chosen randomly. Since this leads to strong overlaps of the particle interaction radii $R$, we let the system first evolve in time with repulsive interactions only, until most of of the overlaps vanished and the defect ratio has reached a value of $d=0.2$. Then, both propulsion and alignment interactions were switched on, and data were recorded.
\vspace{0.1cm}\\
\emph{($\mathcal{H}$) Hexagonal.}
Particle orientations were chosen randomly, while their positions were placed in a perfect hexagonal configuration with a hexagonal lattice spacing of $2R$. This implies a global hexatic order parameter $\Psi_6=1$ and the absence of defects, \emph{i.e. $d=0$}, at the time when simulations were started. \vspace{0.1cm}\\

\noindent Initial conditions used for figures in the main text:\\
\noindent Fig.~1: $\mathcal{R}$;  
\\Fig.~2: $\mathcal{R}$;\\
Fig.~3(a), (b) [circle,triangle], (e): $\mathcal{R}$ and $\mathcal{H}$ lead to equivalent results; 
\\Fig.~3(c) [square], (d): $\mathcal{H}$.

\subsection{Triangulation}

We used standard 2D Voronoi triangulation functions as implemented by the \emph{CGAL} library [http://www.cgal.org]. 

\subsection{Fourier transformation and structure factor}

In Fig.~3 we computed the pair correlation function $g(\vec r) = \left(L/N\right)^2\sum_{i,j} \delta \left( \vec r - (\vec r_i-\vec r_j) \right)$, and the corresponding static structure factor  $S(\vec q)=1 + \frac{N}{L^2} \int \text{d}r^2 g(\vec r) \text{e}^{i\vec q\cdot \vec r}$.  We numerically determined $S(\vec q)$ by using standard fast Fourier transform (FFTW) libraries with a  spatial resolution of $0.1$ unit length.

\subsection{Correlation function of the translational order parameter}

Since global (bond) orientational order is not perfect for $\nu>1$ (\emph{e.g.}\ for $\rho=0.85$, $\Psi_6 \approx 0.9$),  the inverse lattice vector $\vec G$ must be determined properly. We varied the corresponding angle of the inverse lattice vector $\vec G$ with a step size of $5 \cdot 10^{-4}$ rad, and thereby determined the optimal value of $\vec G$ with the weakest decay of $C_{\vec G}$; the ensuing results are depicted in Fig.~3(d,e,g,h), and Fig.~S5 and Fig.~S6 in the Supplementary Material. Note that a non-optimal choice of the lattice angle leads---even for a perfectly hexagonal configuration---to an underestimation of translational order~\cite{underestimation_of_trans_order}.

\subsection{Video descriptions}

\noindent For all attached videos we chose the following  parameter values: Interaction radii   $R= 0.5$, 
updating time    $dt = 1.0$,
alignment strength   $v_a = 0.05$, and
system size     $L = 100$. \\

\noindent The videos depict the \emph{Voronoi triangulation} of the particles (left) and the 
 \emph{local hexatic order parameter} $\left|\Psi_{6,i}\right|$ (right).\\
\noindent  \underline{Voronoi triangulation:} Particles with a coordination different from $6$-fold are illustrated by color: red$=5$ neighbors, yellow$=7$ neighbors, green is equal to more than $7$ or less than $5$ neighbors. \\
\noindent \underline{Local hexatic order parameter:}
 1 = yellow, 0 = black, with the color code shown in Fig.~1 (main text).\\

\noindent Parameters for density $\rho$ and $\vrva$ are indicated in the file names. 
Each video corresponds to an initialization with different \emph{random} particle coordinates and orientations ($\mathcal{R}$).

\cleardoublepage
\newpage

\subsection{Additional Material}

In this section we provide additional graphs 
substantiating certain statements made in the main text.

%
%
\begin{figure}[b]
\includegraphics[width=0.5\linewidth]{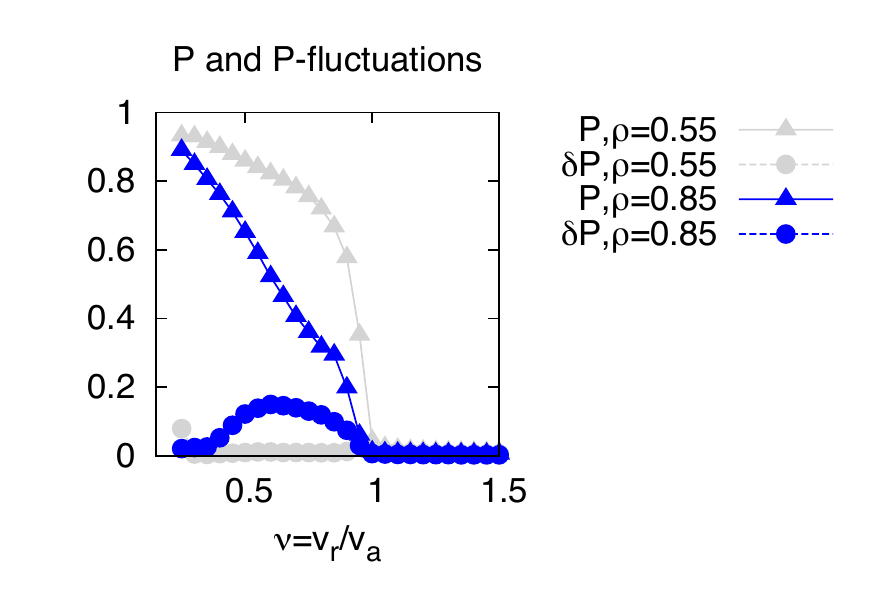}
\caption{\label{fig:varP} 
\emph{Intermittent states:} In oder to find  the $\nu$-parameter regime for the \emph{intermittent states}, we analyzed the total polarity $P$ and the fluctuations in time of $\mathcal{P}(t)$, denoted as $\delta P$. Both quantities are depicted as a function of $\nu$ for a \emph{fluid-like} density $\rho=0.55$ and a \emph{crystalline} density $\rho=0.85$ [refer to Fig.~1, main text].
Pronounced fluctuations in polarity $\mathcal{P}(t)$  exist for $0.375 \lesssim \nu \lesssim 1.0$ (\emph{i.e.}\ $\delta P>0.08$), indicating the \emph{intermittent} regime.
Moreover, our results indicate that the unpolarized--polarized transition with a \emph{fluid-like} density $\rho=0.55$ is  very steep, reminiscent of the \emph{discontinuous} phase transition found in the Vicsek model without repulsion~\cite{Chate:2008}.  In contrast, the existence of the intermittent states at large densities (\emph{e.g.}\ $\rho=0.85$)
flattens the slope of the total polarity in the transitional region between the \emph{crystalline} regime ($\nu>1$) and the polarized non-intermittent regime ($\nu\lesssim0.375$). 
}
\end{figure}

\begin{figure}[tb]
\includegraphics[width=0.7\linewidth]{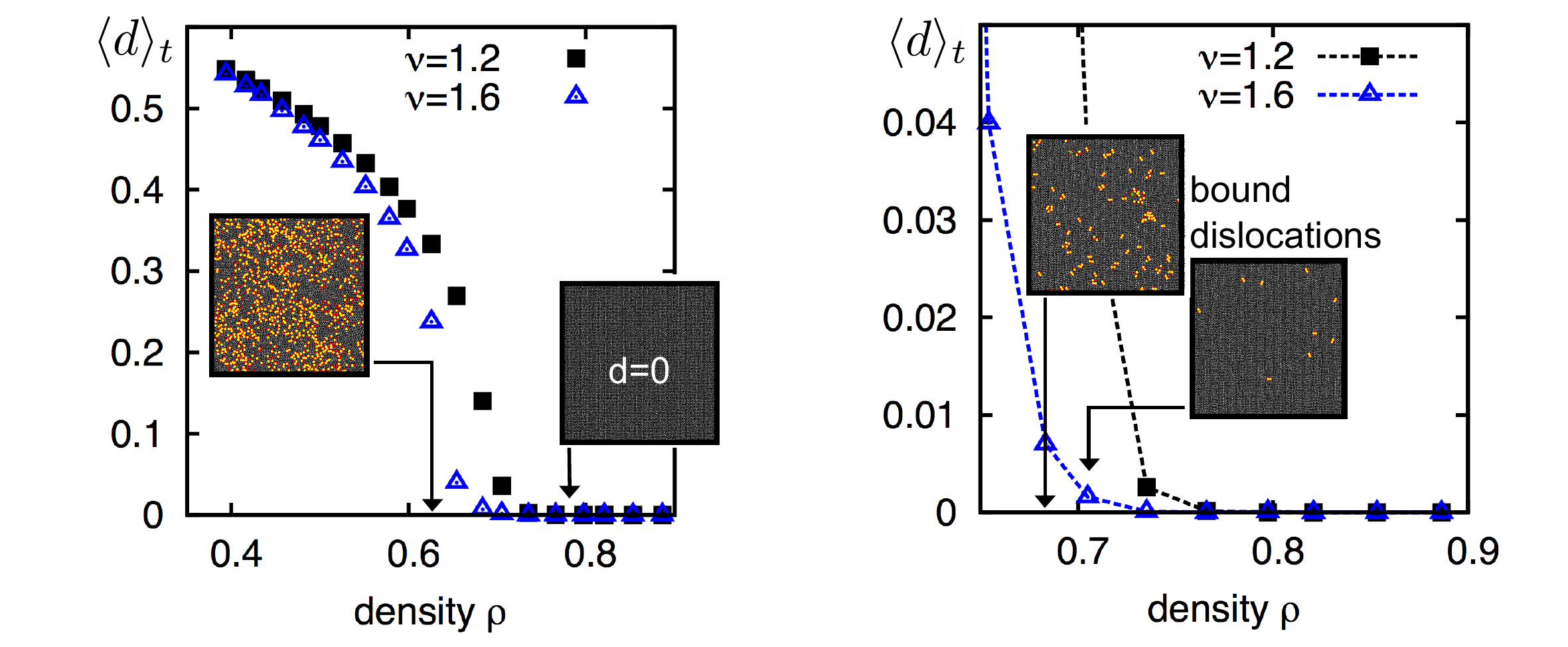}
\caption{\label{fig:defect_phase_transition}  
\emph{Averaged defect ratio $\langle d \rangle_t$} for $\nu=\{1.2, 1.6\}$ as a function of density $\rho$ (\emph{left} and \emph{right} solely differ in a different plot range for the density). The system is initialized in an unpolarized and fully ordered hexagonal configuration ($\mathcal{H}$). The \emph{fluid-like} phase is characterized by a rather large defect ratio $\langle d \rangle_t\sim 0.4$.
Increasing the density, we observe a small density region (\emph{e.g.}\ around $\rho\approx0.7$ for $\nu=1.6$), where $\emph{active}$ fluctuations are strong enough to create a non-zero defect ratio. However, the defect ratio is rather small, \emph{i.e.}\ $\langle d \rangle_t\sim 0.005$. Moreover, within this region, isolated topological defects are absent and 5/7-fold defects are always found in pairs. This indicates that the corresponding states  exhibit \emph{quasi long-range order} and that there is \emph{no hexatic phase} in our model.
Increasing the density further ($\gtrsim0.775$ for $\nu=1.6$), one observes a \emph{defect-free} stationary state with $d=0$.
The decay of the averaged defect ratio can be roughly fitted by $\langle d \rangle_t\propto (\rho_c-\rho)^{\delta}$, with $\delta$ being in the interval $[0.35,0.4]$.
\emph{Snapshots:} Each snapshot depicts the state's triangulation  for a representative configuration in time and corresponds to $\nu=1.6$ (blue data points).  The respective density $\rho$ is indicated by means of the black arrows.
  }
\end{figure}

\begin{figure}
\includegraphics[width=0.35\linewidth]{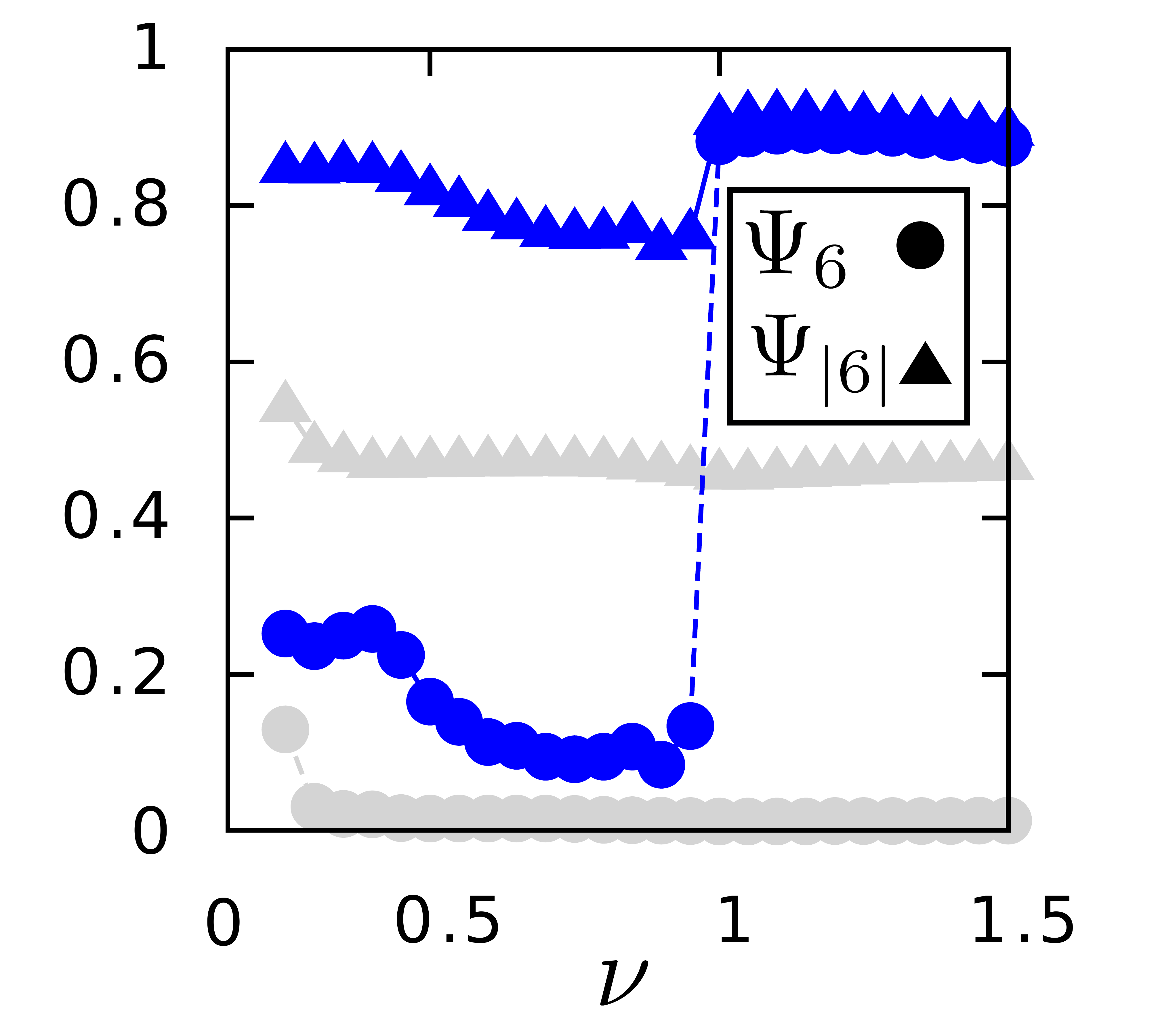}
\caption{\label{fig:hexorderspatialcorr} 
\emph{Local and global hexatic order parameter}, $\Psi_{|6|}$ and $\Psi_{6}$,  as a function of $\nu$ for different values of $\rho$ [$\rho=0.85$ (blue/black); $\rho=0.6$ (light grey)]. Initial conditions: $\mathcal{R}$.\\
  \emph{Interpretation:} To further quantify the nature of the polycrystalline state  we compare the global hexatic order parameter, $\Psi_6=\langle \left|\langle \Psi_{6,i} \rangle_i \right|\rangle_t$, with an order parameter characterizing local hexatic order, $\Psi_{|6|}:= \langle \langle \left| \Psi_{6,i} \right| \rangle_i \rangle_t$. For polycrystalline solids, one expects $\Psi_6$ to be small since the complex numbers $\Psi_{6,i}$ of particles from different ordered patches cancel in the average $\langle \ldots \rangle_i$. In contrast, using the absolute values $\left| \Psi_{6,i} \right|$ there is no such cancellation of phases. Therefore, the value of $\Psi_{|6|}$ should be close to $1$ in a polycrystalline phase because most particles have six-fold coordination and are not located at grain boundaries. The figure above depicts both order parameters as a function of $\nu$ for two different densities, with the lower value corresponding to the \emph{fluid-like} and the larger one to the \emph{crystalline} regime. In the \emph{fluid-like} regime, local  bond orientational order, $\Psi_{|6|}$, is moderately developed while global orientational order, $\Psi_{6}$, is close to zero. Moreover, both order parameters are only weakly dependent on $\nu$. This is in stark contrast to the behavior at higher densities, $\rho=0.85$. There $\Psi_{6}$ shows a steep and large decrease at $\nu\approx1$, while $\Psi_{|6|}$ remains approximately constant, indicating that we have a polycrystalline phase for $\nu \lesssim 1$. Since the active crystal phase ($\nu\gtrsim 1$) exhibits quasi-long-range order, local and global hexatic order are likewise well developed. 
 }
\end{figure}
\begin{figure}[b]
\includegraphics[width=0.65\linewidth]{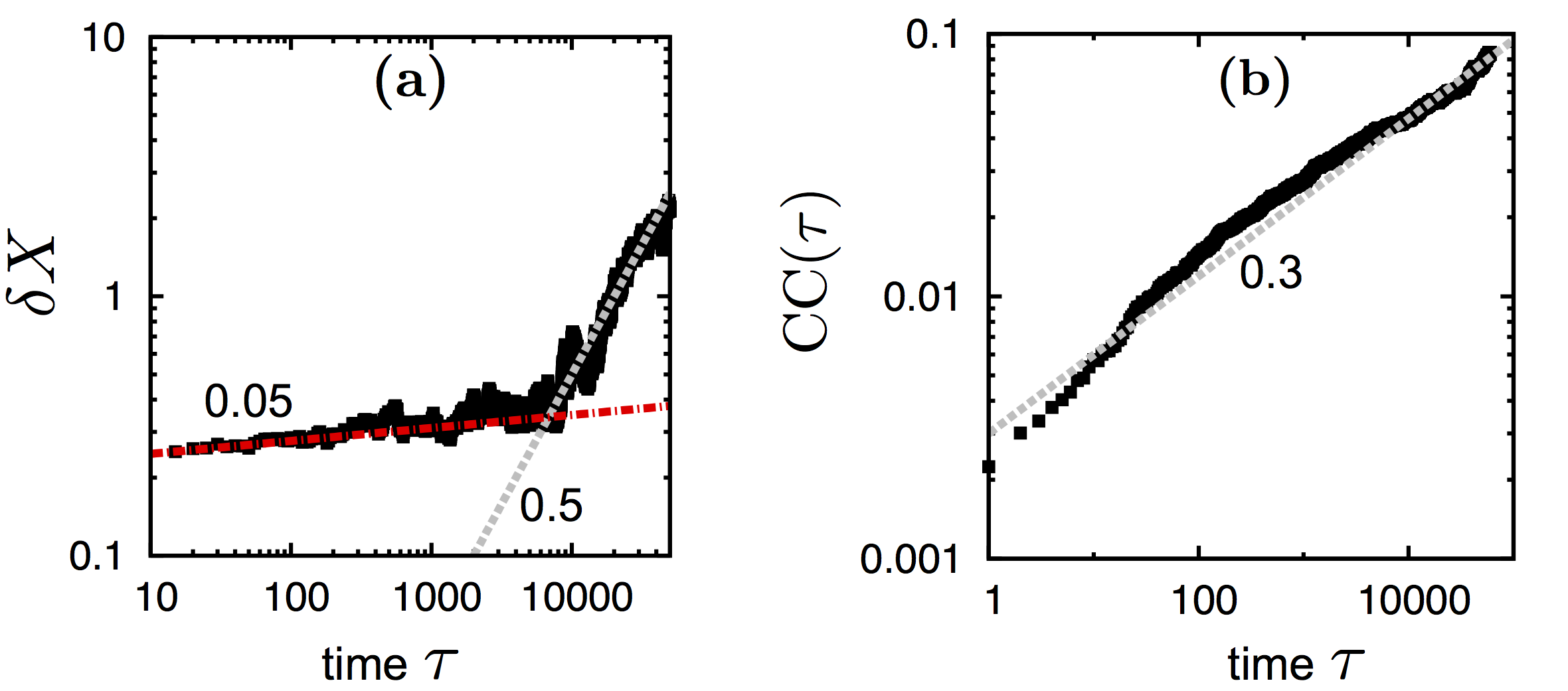}
\caption{\label{fig:msd}  
\emph{Motion of topological defects:} 
Since the number of topological defects is rather small in the stationary state, their motion cannot be characterized by studying the defect's mean square displacement within an appropriate statistical accuracy. 
Therefore, we analyzed \textbf{(a)} the \emph{particles' mean square displacement}, $\delta X(\tau)=\sqrt{\langle [\vec x_i(t_0+\tau)-\vec x_i(t_0)]^2  \rangle_i}$, with $i$ denoting the particle index and $t_0$ is a time point where the number of defects $d(t)$ have become approximately stationary. 
We find a \emph{sub-diffusive} particle motion for small and intermediate time scales with $\delta X(\tau) \sim \tau^{0.05}$ (red dash-dotted line). At large time scales there is a  crossover to a diffusive regime with $\delta X \sim \tau^{0.5}$ (grey dotted line). 
Moreover, we evaluated \textbf{(b)} the \emph{cage correlation function}~\cite{Weeks_Weitz}, $\text{CC}(\tau)$, here defined as the ratio of particles which have changed at least once their neighborhood (often referred to as ``cage"). In the time regime corresponding to sub-diffusive particle motion, mostly none of the particles have rearranged their neighborhood. 
We find that in the considered time regime $\text{CC}(\tau)$ increases according to a power-law $\text{CC}(\tau)\sim\tau^{0.3}$, whereby the ratio of rearranged cages remains relatively small; until the crossover to the diffusive regime occurs, only about $5\%$ of the particles have rearranged their cage(s). \\
Since defects can either move by cage rearrangements or particle motion, we can conclude that defects also move sub-diffusively---as the particles---at least for small time scales. 
The slow movements of the topological defects makes it practically unfeasible to follow numerically the process $d \to 0$ for large time scales when starting with a random initial condition ($\mathcal{R}$), \emph{i.e.}\ the annihilation processes of just a few remaining isolated topological defects in the system.\\
Parameters: $\rho=0.75$, $\nu=1.2$, random initial condition $\mathcal{R}$.  
 }
\end{figure}
%
%
%
%
\begin{figure}[b]
\includegraphics[width=0.5\linewidth]{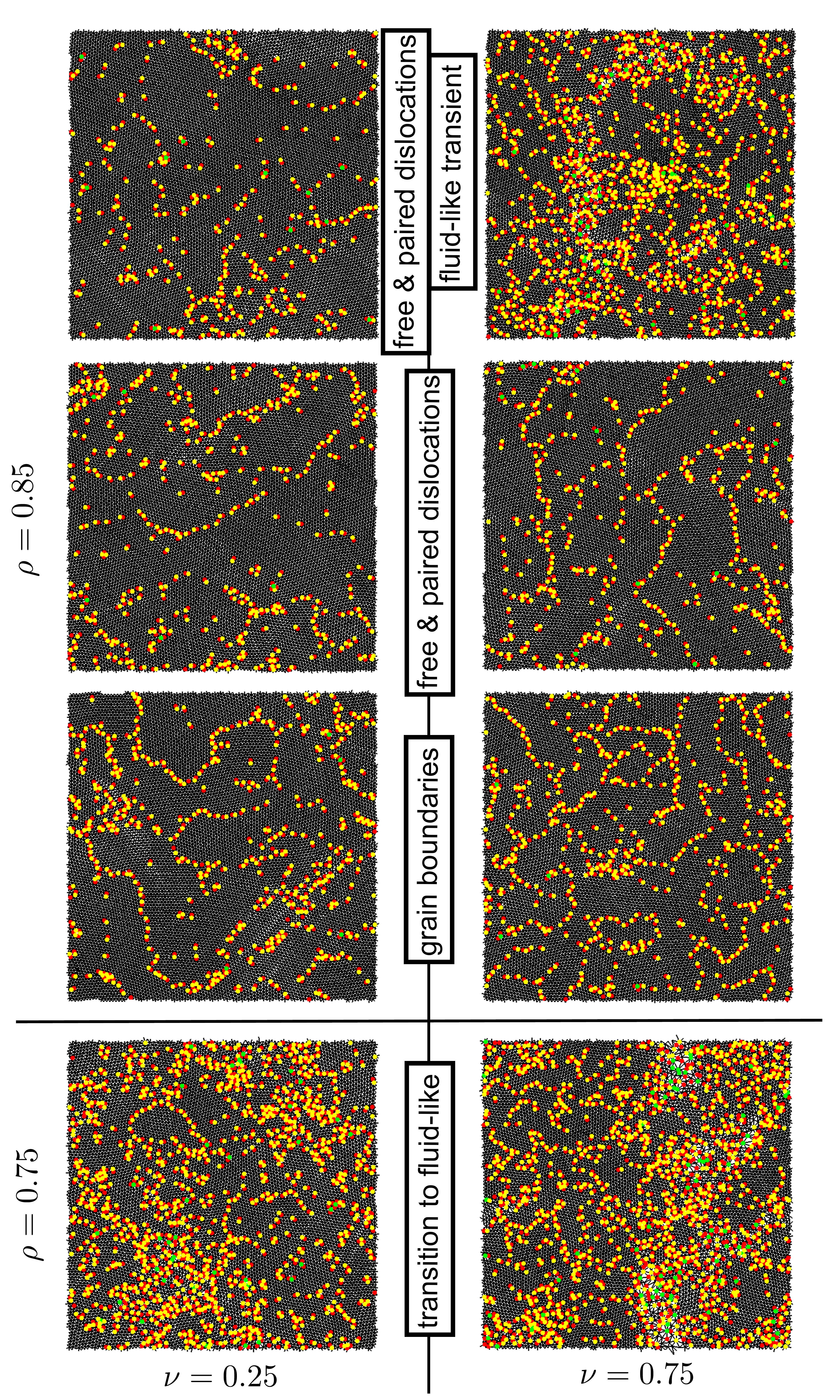}
\caption{\label{fig:Spat_arrangement_of_defects}  
\emph{Illustration of the spatial arrangements of grain boundaries and defects} for $\nu<1$ (\emph{left:} weakly intermittent $\nu=0.25$; \emph{right:} intermittent state $\nu=0.75$), and densities $\rho=0.85$ and $\rho=0.75$. For $\rho=0.85$, three representative configurations in time (no particular order in time) 
 are depicted for each of the two $\nu$-values.
Even though intermittency is different for $\nu=0.25$ and $\nu=0.75$, one finds a similar arrangement of topological defects ranging from periods with mostly connected grain boundaries, to periods of pronounced coexistence between paired and free 5/7-fold defects. In contrast, for $\rho=0.75$,  defects are large in number  ($d\sim0.2$) and appear spatially disordered. Therefore, we term this state fluid-like.
}
\end{figure}
%
%
\begin{figure}[b]
\includegraphics[width=0.8\linewidth]{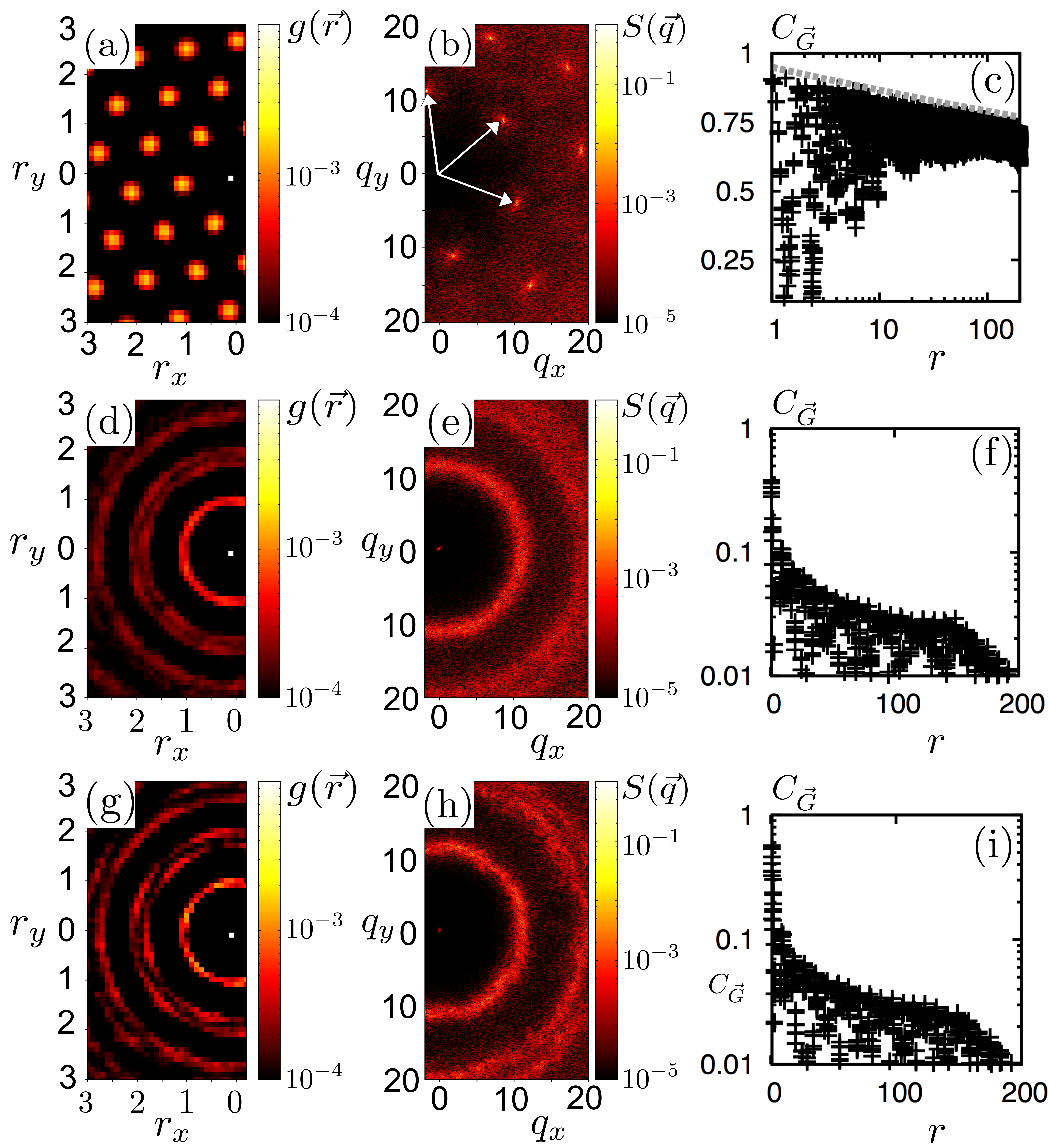}
\caption{\label{fig:spatorder_intermittent} 
\emph{Pair correlation} function $g(\vec{r})$, \emph{static structure factor} $S(\vec q)$ and \emph{Correlation function} $C_\vec{G}$ for decreasing values of $\nu$ (from top to bottom). \textbf{(a,b,c)} corresponds to
 $(\rho,\nu)=(0.85,1.5)$, \textbf{(d,e,f)} to $(\rho,\nu)=(0.85, 0.75)$ and \textbf{(g,h,i)} to $(\rho,\nu)=(0.85, 0.25)$, respectively. Reciprocal lattice vectors $\vec G$ are indicated by white arrows in \textbf{(b)}.
Note that since the stationary states for $\nu <1$ develop very fast, there is no difference in the results obtained by either starting from a hexagonal ($\mathcal{H}$) or a disordered ($\mathcal{R}$) configuration. Results correspond to a simulation box of size $L=400$ containing $N=172156$ particles. }
\end{figure}
%
%

\begin{figure}[b]
\includegraphics[width=0.8\linewidth]{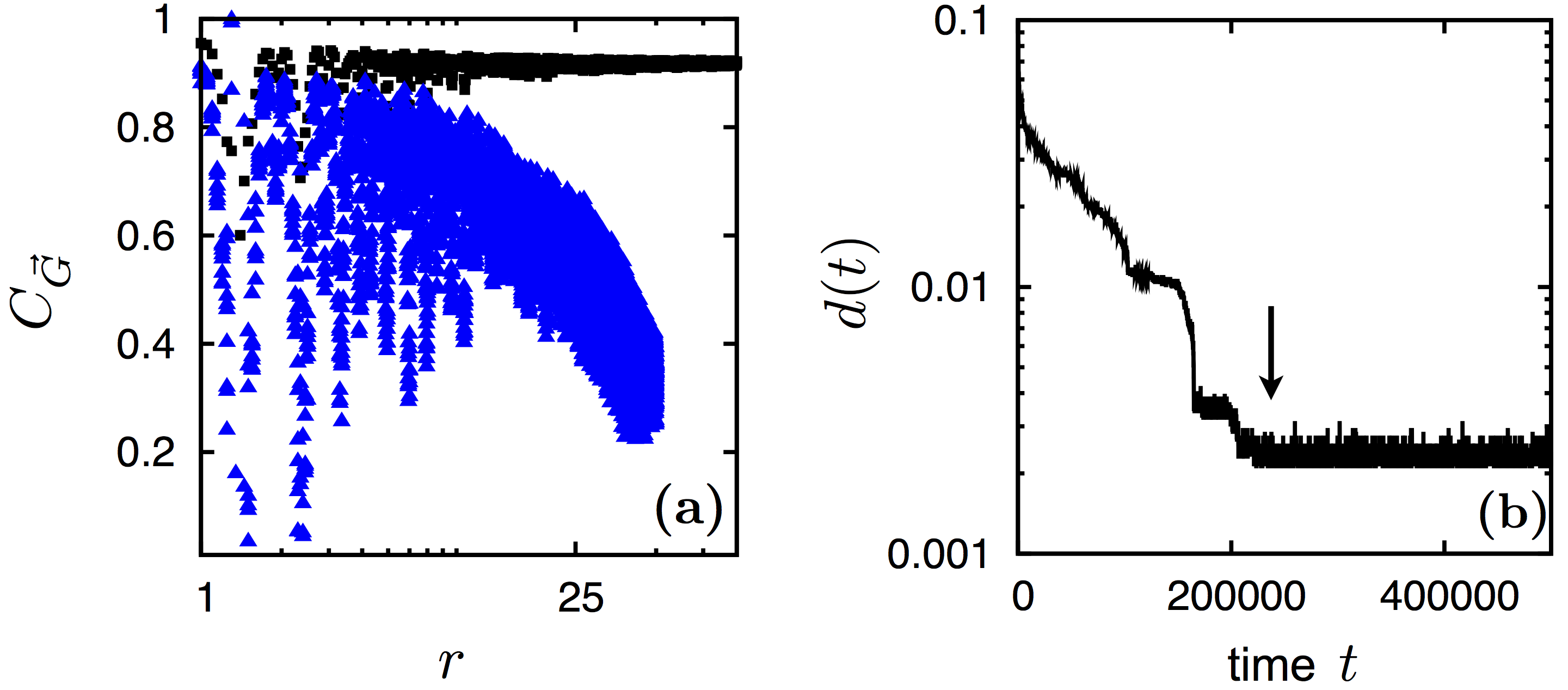}
\caption{\label{fig:C_G_comparision_init_plus_time_trace} 
\textbf{(a)} Comparison of the translational correlation function $C_\vec{G}$ of a considered \emph{crystalline state} (Parameters: $\rho=0.85$, $\nu=1.5$) for the \emph{two} different initial conditions: \emph{Random}($\mathcal{R}$) and \emph{hexagonal}($\mathcal{H}$). \\
The black curve in \textbf{(a)} depicts the results already presented in the main text [Fig.~3(d)]. It corresponds to a defect-free ($d=0$) stationary state obtained by using  \emph{hexagonal}($\mathcal{H}$) initial conditions.  The reason for the choice of  \emph{hexagonal}($\mathcal{H}$) initial conditions is that the time for all defects to vanish out of the system increases to time-scales that cannot be addressed numerically (explanation see caption of Supplemental Fig.~\ref{fig:msd}).\\
The blue curve in \textbf{(a)} corresponds to the same parameter set as the black curve, but the system was initialized in a \emph{random} configuration ($\mathcal{R}$). 
For the comparison we selected a realization with the smallest number of topological defects $d$ and computed $C_\vec{G}$ within a ``quasi-stationary" regime, where the defect ratio $d(t)\approx0.0025$. The corresponding time trace of the defect ratio $d(t)$ is shown in \textbf{(b)}. The begin of the``quasi-stationary" regime is marked by an arrow. 
The existence of defects ($d\not=0$) leads to an underestimation of the decay of $C_\vec{G}$, which is depicted in \textbf{(a)} by the blue triangles.
}
\end{figure}

\end{document}